\documentclass[12pt,preprint]{aastex}

\slugcomment{Accepted for publication in {\it{ApJ}} 3 September 2010}

\shorttitle{Star Formation Rates in GMCs}
\shortauthors{Lada, Lombardi \& Alves}

\def\msun{M$_\odot$}
\def\msunsp{M$_\odot$ \ }
\def\cc{cm$^{-3}$}

\def\13co{$^{13}$CO}

\begin{document}


\title{On the Star Formation Rates in Molecular Clouds}


\author{Charles J. Lada }
\affil{Harvard-Smithsonian Center for Astrophysics, 60 Garden Street
Cambridge, MA 02138, USA}
\email{clada@cfa.harvard.edu}

\author{Marco Lombardi}
\affil{European Southern Observatory, Karl-Schwarzschild-Strasse 2, 85748, Garching Germany}
\email{mlombard@eso.org}

\and

\author{Jo\~ao F. Alves}
\affil{Institute for Astronomy,  University of Vienna, T\"urkenschanzstrasse 17, 1180 Vienna, Austria}
\email{joao.alves@univie.ac.at}

\begin{abstract}
In this paper we investigate the level of star formation activity within nearby molecular clouds. We employ a uniform set of infrared extinction maps to provide accurate assessments of cloud mass and structure and compare these with inventories of young stellar objects within the clouds. We present evidence indicating that both the yield and rate of star formation can vary considerably in local clouds, independent of their mass and size. We find that the surface density structure of such clouds appears to be important in controlling both these factors. In particular, we find that the star formation rate (SFR) in molecular clouds is  linearly proportional to the cloud mass ($M_{0.8}$)  above an extinction threshold of A$_K  \approx $ 0.8 magnitudes, corresponding to a gas surface density threshold of  $\Sigma_{gas} \approx 116$ \msunsp pc$^{-2}$. We argue that this surface density threshold corresponds to a gas volume density threshold which we estimate to be n(H$_2$) $\approx$ 10$^4$  \cc. Specifically we find SFR (\msun yr$^{-1}$) = 4.6 $\pm$ 2.6  $\times$ 10$^{-8}$ $M_{0.8}$ (\msun) for the clouds in our sample. This relation between the rate of star formation and the amount of dense gas in molecular clouds appears to be in excellent agreement with previous observations of both galactic and extragalactic star forming activity. It is likely the underlying physical relationship or empirical law that most directly connects star formation activity with interstellar gas over many spatial scales within and between individual galaxies. These results suggest that the key to obtaining a predictive understanding of the star formation rates in molecular clouds and galaxies is to understand those physical factors which give rise to the dense components of these clouds.

\end{abstract}



\section{Introduction}

 The recent recognition of massive nearby clouds with little star formation activity indicates that the star formation rates in even nearby clouds of similar mass can vary considerably, as much as an order of magnitude (Lada et al 2009). Therefore, systematic and comparative observational studies of  the physical properties of local molecular clouds and the relation of these properties to the varying levels of star formation activity within them could lead to new insights concerning the underlying physics controlling the star formation rates in molecular gas.

Over the past few years we have conducted systematic, wide-field, extinction mapping surveys of a significant fraction of the molecular clouds within 0.5 kpc of the sun. One of the most interesting findings of these surveys was the identification of two massive, nearby molecular clouds (The Pipe Nebula and the California Molecular Cloud) with relatively little star formation activity. The existence of such clouds suggests that large (order of magnitude) variations in the levels of star formation activity and thus star formation rates occur in the closest molecular clouds. For example, consider two of the nearest molecular clouds to the sun, the Ophiuchus cloud at 120 pc and the Pipe Nebula at 130 pc. These two dark clouds are close neighbors on the celestial sphere,  separated in projection by 10 degrees (22 pc). They have similar spatial extents ($\approx$ 15 pc) and their total masses are similar to within a factor of two ($\approx$ 14,000 and 8,000 \msun, respectively).  The Ophiuchus cloud, on the one hand,  has been long known for intense star formation activity, it is the birth site an embedded cluster of about 300 stars (e.g., Wilking, Gagne \& Allen 2008 ). The Pipe Nebula, on the other hand,  has negligible star forming activity, containing only about 20 recently formed stars (Forbirch et al. 2009).  Despite their similarity in size and mass, these two clouds are characterized by  rates of star formation that differ by  more than an order of magnitude. Consider also the California Molecular Cloud.  With a mass of about 10$^5$ \msun,  it is considerably more massive that either the Pipe or Ophiuchus clouds. Indeed, it is the most massive GMC within 0.5 kpc of the sun and is similar in size, mass and distance to the much better known Orion A GMC. However the star formation rate in the California cloud is only comparable to that in the Ophiuchus cloud and an order of magnitude lower than that in the Orion A cloud (Lada et al. 2009). 

Because of their large differences in star formation rates and their proximity to the sun, these clouds provide excellent laboratories for investigation of the physical factors that control the rate of conversion of molecular gas and dust into stars. Indeed, Lada et al. (2009) reported the intriguing results that the differences in star formation rates between the California,  Orion and Pipe molecular clouds were reflected in similar differences in the amounts of the high extinction (i.e., A$_k$ $\geq$ 1.0 magnitudes), and presumably dense (n(H$_2$) $\geq$ 10$^4$ \cc) material contained in these clouds and that this could indicate a density threshold for star formation. These results could represent a potentially significant clue concerning the physical origin of the different levels of star formation activity in these nearby clouds. To explore the relationship between star formation and cloud structure and to test the hypothesis that the star formation rate is directly related to the amount {\it dense} gas within a cloud, we systematically examine in this paper the physical conditions and star formation activity in a larger sample of nearby clouds for which we have previously obtained wide-field extinction maps.   We explore the relation between cloud structure and star formation in eleven local molecular clouds. We find compelling evidence for large variations in the star formation rates between these clouds. Furthermore, our data are consistent with the existence of  a column density threshold for star formation activity above which the star formation rate appears to be linearly correlated with total cloud mass above the threshold. We estimate this threshold to be at an extinction of approximately 0.8 $\pm$ 0.2  magnitudes at 2.2 $\mu$m, corresponding to a total gas column density of about 116 $\pm$ 28 \msun pc$^{-2}$ and a cloud volume density of roughly  10$^4$ cm$^{-3}$.  Below this threshold star formation activity is presumably negligible.

\section{Observations}

\subsection{The Molecular Cloud Sample}

Over the past few years we have used the 2MASS infrared sky survey to systematically construct and analyze wide-field extinction maps of  prominent dark clouds in the solar vicinity.  To date this survey program has produced maps of eight molecular  cloud complexes within 450 pc of the sun. These are the Pipe Nebula (Lombardi et al. 2006), the Ophiuchus cloud and the Lupus cloud complex (Lombardi et al. 2008), the Taurus, Perseus and California clouds (Lombardi et al 2010; Lada et al. 2009), the RCrA cloud (Alves, Lombardi \& Lada 2010, in preparation) and the Orion cloud complex (Lombardi, Lada \& Alves 2010, in preparation). We have used these maps to measure the masses and investigate in some detail the structure of  these nearby clouds. The maps were primarily made using the NICER method (Lombardi \& Alves 2001), an optimized technique for converting the individual multi-band photometric measurements of millions of stars into well ordered, fully sampled, extinction maps of large areas of the sky at modest angular resolutions. The angular resolutions employed were individually optimized to the various clouds studied to obtain the highest spatial resolution possible given the distance to the cloud and the varying surface densities of background and foreground stars.  These maps have provided accurate and robust measurements of the total masses of these clouds as well as information concerning their structural properties. 

In this paper we are interested in comparing the structure of clouds at relatively high extinctions (A$_ K$ $>$ 0.2 mag). Measuring cloud masses at the highest extinctions (A$_K \approx$ 1 mag) can be very sensitive to the angular resolution employed and thus both to the distance of a cloud and the surface density of field stars in its direction. At high extinction levels systematic, unresolved  extinction gradients, random small scale structure and foreground star contamination at sub-pixel scales, can bias extinctions measured with the NICER technique to lower values. For this reason we have re-measured all the clouds in our sample using the NICEST algorithm  developed by Lombardi (2009) to minimize these effects. In Table 1 we list the clouds investigated in this paper along with the angular resolutions employed in the NICEST maps and the distances we adopted for each of the clouds.

In Figure 1 we present the cumulative mass profiles of the clouds as a function of infrared extinction (A$_K$) derived from the NICEST measurements. The profiles show interesting behavior. They all fall with increasing extinction in a non-parallel fashion and appear to diverge at high extinctions.  These profiles are similar to ones published in our earlier papers ( e.g., Lombardi et al. 2006, 2008, 2010; Lada et al. 2009), however  some  differences exist between the profiles presented here and the earlier work and these differences are the result of both the use of NICEST and the different resolutions employed in the respective analyses. The profiles in Figure 1 are also generally similar to those recently published by Kainulainen, et al. (2009) and Froebrich \& Rowles (2010). We also list the cumulative masses of the clouds in our sample at two different values of extinction in Table 2. The first, at A$_K$ $=$ 0.1 magnitudes, corresponds to our estimate for the total mass of the cloud and the second, at A$_K$ $=$ 0.8 magnitudes, corresponds to our estimate for the mass of the dense gas component of the cloud.

\subsection{Inventory of Star Formation Activity}

In the past few years ground-based optical and infrared and space-based infrared surveys of nearby clouds have provided a wealth of data on the populations of embedded Young Stellar Objects (YSOs) in nearby molecular clouds.  For each of the clouds we conducted a survey of the recent literature and compiled a list of their total YSO population. The results are listed in Table 2 along with the corresponding references used to estimate the sizes of the various cloud YSO populations. Unfortunately not all the clouds have been surveyed completely or thoroughly to similar sensitivities. Where possible we used an infrared census to estimate the size of the associated YSO population in order to insure some minimal level of uniformity. Fortunately most of these clouds have been observed with the Spitzer Space Telescope and for many of these fluxes and positions are publically available on the Web (i.e., C2D survey: http://irsa.ipac.caltech.edu/data/SPITZER/C2D/). Many of the clouds have also been observed with ground-based near-infrared imaging surveys. The most complete inventory exists for the Pipe Nebula, followed by Taurus, Ophiuchus and Perseus, then Orion A, and B and finally the California Molecular Cloud. However, in no case do we expect the inventories to be off by a factor of more than 2, at the most. If there is a bias it is that infrared surveys tend to not be complete for Class III sources since lacking strong infrared excesses they can be undercounted. This is not likely the case for the Perseus,  Taurus, Lupus and RCrA clouds whose Class III populations are  well represented in existing tabulations of membership. The situation for the other clouds is less clear in this regard. We also count only the YSOs or candidate YSOs that are within the cloud boundaries (i.e., within the A$_K$ = 0.1 magnitude contour). Typically this selection includes the vast majority of known members in a cloud and our census represents a more or less complete inventory of star formation activity over the last 2 million years or so in each cloud (see below). 


For the purposes of this paper we will assume that the size of a YSO population is directly related to the star formation rate (SFR) in a cloud. This is likely a good assumption given that when plotted on HR diagrams the YSOs in these clouds all seem to have similar ages and ages spreads with a median age of 2 $\pm$ 1 million years (e.g.,  Covey et al. 2010 and many others). The median mass for the IMF of a stellar population is 0.5 \msunsp (e.g., Muench et al. 2007).  Thus we derive a SFR for each cloud from:

 \begin{equation}
SFR = 0.25 N(YSOs)\  \times 10^{-6}  \ \  M_\odot \  yr^{-1} 
\end{equation}

\noindent
and list the rate for each cloud in Table 2.

\section{Results and Analysis}

\subsection{On The Variation of Star Formation Activity and Rates in Molecular Clouds}

Examination of the data we have compiled  (e.g., Table 2) provides evidence for significant variations in star formation activity and star formation rates  among local molecular clouds.  This is shown in Figure 2 where we plot the ratio of N(YSO), the size of the YSO population, to $M_{tot}$, the total mass of the cloud (i.e., $ M_{tot} = \int^{+\infty}_{0.1} M(A_K) dA_K$) as a function of the total cloud mass. This ratio is both a measure of the star formation efficiency in the clouds and, from   equation 1, also  a measure of the star formation rate per unit cloud mass. As is clearly seen in the plot, the star formation efficiency (and SFR per unit mass) at a given cloud mass varies considerably, by well more than an order of magnitude, between the local clouds. 

 Evidence for significant variations in the star forming activity of at least some molecular clouds is not new. The GMC known as Maddalena's cloud  has long been known as an example of a massive GMC without significant star formation ( Maddalena \& Thaddeus 1985; Lee et al. 1994; Williams and Blitz 1998; Megeath et al. 2009). However the lack of star formation in this  massive cloud was considered to be a rare phenomenon and the nature of the cloud as either an extremely young GMC (Maddalena \& Thaddeus 1985) or an old relic of some earlier episode of star formation (Lee et al. 1994) has been the subject of debate. However, the recent recognition of the nearby, massive California molecular cloud as a GMC with low levels of star formation activity coupled with the results presented here in Figure 2, suggest that significant variations in star formation activity may not be that rare. Indeed, in a paper published more than twenty years ago, Mooney and Solomon (1988) combined observations of $^{12}$CO with IRAS infrared images of a sample of 55 clouds in the inner Galaxy and presented evidence for the existence of very large variations in the star formation rates per unit mass among these inner Galaxy GMCs. In a plot similar to Figure 2 here they found variations of roughly two orders of magnitude in the star formation rates per unit cloud mass for clouds covering a mass range of roughly 10$^4$ to 10$^6$ \msun. 
 
 Our results more securely and firmly establish that significant variations in the level of star formation activity and star formation rate are common in Galactic molecular clouds and are largely independent of cloud mass.  As a result, it should be possible from detailed comparative studies of such clouds to determine the physical factors that govern the star formation rates in molecular gas.

\subsection{The Relation Between Cloud Mass and Star Formation}

One would naturally expect there to be a general correlation between the quantities N(YSOs) and M$_{tot}$. The most massive clouds should produce more stars than the least massive clouds. However, as can be ascertained from Figure 2, this correlation will not be particularly interesting in any reasonable predictive sense because of the extremely large variation in star formation efficiencies between the clouds. For example, consider the first three entries in Table 2, the three most massive clouds in our sample. Here the correlation is reversed from what one would expect, for in this group the more massive the cloud the lower its YSO content.

One might expect a tighter correlation to exist  between star formation activity and the amount of high extinction material in a cloud. This is because it has been known for some time (i.e., Lada 1992) that active star formation is confined to the high volume density regions of molecular clouds. Indeed, Lada, Evans and Falgarone (1997) found star formation efficiencies, and presumably star formation rates, to be higher in the massive dense cores of the L1640 (Ori B) cloud that had the largest amounts of gas at densities of n(H$_2$) $\approx$ 10$^5$ \cc.  Since molecular clouds are stratified with the highest extinction material typically confined to narrow (0.2 - 0.3 pc) filamentary structures and the inner regions of dense cores, high extinction material typically corresponds  to  high volume density material. 
To investigate the possibility of a more meaningful correlation between star formation and high extinction molecular gas we plot in Figure 3 the ratio $M_{A_{K0}}/N(YSO)$ as a function of extinction, $A_K$, where $M_{A_{K0}}$ is the cumulative mass above a given threshold extinction, $A_{K0}$ (i.e., $M_{A_{K0}} = \int_{A_{K0}}^{+\infty}$ $M(A_K) dA_K$). The curves generally decline with extinction as would be expected, but in addition they do appear characterized by a dispersion which has a minimum near  A$_K$ $\approx$ 0.8 magnitudes. The presence of such a minimum in the dispersion of these curves does indeed suggest that a stronger correlation between star formation activity and cloud mass exists for  higher extinction material. For example, in Table 2 we list the cumulative masses of the clouds above an extinction threshold of  A$_{K0}$ $=$ 0.8 magnitudes. Although not perfect, the ordering of these masses does correspond more closely to the ordering of the sizes of the YSO populations than do the total cloud masses.   

To make a more quantitative assessment of this apparent minimum in the dispersion of the cumulative mass profiles we caculated the dispersion of the curves in Figure 3 by dividing the  standard deviation of the  curves as a function of extinction by the mean value at that extinction.  The so calculated dispersion is shown as a function of extinction in the bottom panel of Figure 3. We also calculated the dispersion in the logarithms of the curves and plot that as a dashed trace in the lower panel of Figure 3.
The calculation confirms the presence of a broad minimum in the normalized dispersion between roughly 0.6 and 1.0 magnitudes of extinction. The dispersion is about a factor of 2-3 lower in this extinction range than it is at the lowest extinctions. The existence of such a minimum indicates that there is an extinction or range in extinction at which the ordering of cloud mass (above that extinction) reflects most directly the ordering of the SFR that characterizes the clouds. This suggests that a relation of the form $N_{YSOs} = M^\alpha_{A_{K0}}$ could exist near the minimum of the dispersion with $\alpha \approx 1$. To further investigate this possibility we performed least-square fits to a series of plots of the quantities $N_{YSOs} $ vs $M_{A_{K0}}$ to determine the value of $A_{K0}$ at which $\alpha$ was closest to 1.0. This would be the value of the extinction at which the cumulative masses of the clouds most directly reflected the star formation activity within them. We estimated this extinction to be $A_{K0}$ $=$ 0.8 $\pm$ 0.2 magnitudes, corresponding to A$_V$ $=$ 7.3 $\pm$ 1.8 magnitudes and $\Sigma_{gas} =$116 $\pm$ 28 \msun pc$^{-2}$. This was also the value of extinction that produced the smallest uncertainty in the fits to the two quantities. Figure 4 is a plot of  N$_{YSOs}$  vs $M_{0.8}$, the integrated cloud mass above the extinction of  0.8 magnitudes. The two observed quantities indeed appear linearly correlated and the least-squares fit to this data produced a slope, $\alpha$ = 0.96 $\pm$ 0.11.

We note here that for this analysis we have compared the total YSO yield of the various clouds to their masses above the extinction threshold of 0.8 infrared magnitudes. Many of the YSOs we counted may have evolved and migrated away from their birthplaces or dispersed much of their parental material. Such evolved young stars are likely to be located in regions of the cloud where the extinction is less than 0.8 magnitudes and the densities lower. One might expect an improved, if not more appropriate, comparison would result from only counting those YSOs at higher extinction levels.  We can only examine this possibility for those clouds in our sample where there are both statistically significant samples of YSOs at high extinctions and published YSO positions  (i.e., Ophiuchus, Pipe, Taurus, Perseus, Lupus 3, 4, RCrA). Indeed, counting only YSOs at A$_K$ $\geq$ 0.5 magnitudes in these clouds, we again found a strong, linear correlation between YSO number and threshold mass, with a least-squares derived slope of 0.96 $\pm$ 0.13, essentially identical to that for the lower extinction threshold sample. That this refined sample gives the same result as the entire sample suggests that any gas lost in the star formation process, and thus not counted in our cloud mass measurements, has little effect on the overall result. Apparently star formation activity over the last 2 Myrs has not produced a significant modification to the total mass of high extinction material contained within the clouds in our sample.

The  linear nature of the correlation in Figure 4 indicates that it is the cloud material above the  extinction of approximately 0.8 magnitudes that is most directly related to the level of star formation activity and the SFR in a cloud. These results also suggest that there is a threshold extinction above which the number of YSOs produced by a cloud (and the SFR ) is directly proportional to the mass of the cloud above that threshold.    We can now describe the relation between the SFR and cloud extinction as follows:

\begin{equation}
 SFR(A_{K})  = \cases {  0 & A$_{K} <$ A$_{K0}$ $\approx$ 0.8 mag; \cr \epsilon  M_{A_{K0}} / \tau_{sf} & A$_{K} \geq$ A$_{K0}$ $\approx$ 0.8 mag. \cr}\end{equation}
 
\noindent
where $M_{A_{K0}} = \int_{A_{K0}}^{+\infty}$ $M(A_K) dA_K$, $\tau_{sf}$ is the timescale of star formation and $\epsilon$ is the present star formation efficiency in the gas of mass $M_{A_{K0}}$. However, we note here that the present data suggest that the extinction threshold of 0.8 magnitudes  may not be particularly sharp, spanning a range of 0.6 to 1.0 magnitudes.  Nonetheless using the data in Table 1 for A$_{K0}$ $=$ 0.8 magnitudes and assuming $\tau_{sf}$ $= $ 2 $\times$ 10$^6$ yr$^{-1}$, we find that $\epsilon =$ 10 $\pm$ 6 \%. We can further express the star formation timescale in terms of a free-fall timescale at the density corresponding to the threshold extinction: $\tau_{sf} = f \times  \tau_{ff}$;
where $f \geq 1.0$,  $\tau_{ff} = ({3\pi / 32G\rho_t})^{0.5}$ and $\rho_t$ is the density at
the threshold extinction. Then above the extinction threshold, 

\begin{equation}
SFR(M_{A_{K0}}) = {\epsilon \over f}   \times  \tau_{ff}^{-1}\vert_{\rho_t} \times M_{A_{K0}}
\end{equation}

\noindent
where $\tau_{ff}$ is evaluated at the threshold density. The ratio, ${\epsilon \over f}$, is the star formation efficiency per free-fall time.  For $\rho_t$ $=$ 10$^4$ \cc (see discussion below)  $\tau_{ff}$ $=$ 3.5 $\times$ 10$^5$ yrs, $f$ $=$ 5.7 and the average efficiency per free-fall time is: ${\epsilon /f}$ $=$ 1.8 \%.

\section{Discussion}

As mentioned earlier, the fact that we find star formation to be most intimately associated with the high extinction material in clouds is hardly surprising since it has been known for some time  that active star formation is confined to the high volume density regions of molecular clouds (e.g., Lada 1992) and that regions of high volume density correspond to regions of high extinction.  Therefore it might be reasonable to assume that equation 2 can be rewritten as:

\begin{equation}
 SFR({\rm n(H}_2))  = \cases { 0& n(H$_2) <$ n$_0$\cr \epsilon  M_{0} / \tau_{sf} & n(H$_2) \geq$ n$_0$ \cr}
 \end{equation}
 
\noindent
where $M_{0} =  \int_{n_0}^{+\infty}$ $M(n) dn$ and $n_0$ is the threshold volume density, corresponding to the extinction threshold of A$_K$ $\approx$ 0.8 magnitudes, and it is implicitly assumed that $M_0 = M_{A_{K0}}$.

The value of n$_0$ that corresponds to the threshold extinction of A$_K$ = 0.8 magnitudes is not  known apriori, but can be estimated from observations. Consider that column density, N,  is related to volume density, n, as N(r) = n(r) $\times$ r. For a stratified cloud with n $\sim$ r$^{-p}$, N $\sim$ r$^{1-p}$. Using this fact Bergin et al. (2001) modeled the observed extinction profile of the IC 5146 (B168) dark cloud assuming cylindrical geometry and p $=$ 2 and determined the relation between A$_V$ and n(H$_2$) for that cloud. They found (c.f. their figure 10)  that a density of n(H$_2$) = 10$^4$ cm$^{-3}$ corresponded to an extinction of A$_V$ = 6 magnitudes (A$_K$ = 0.66 mag.). It has been long known that molecular lines such as  NH$_3$, N$_2$H$^+$, and HCN require relatively high densities (n(H$_2$) $\ga$ a few $\times$ 10$^4$ cm$^{-3}$) to be observed. These species are always detected in regions of high extinction, but direct, quantitative comparison with extinction measurements have been rarely made. However such comparisons do exist for N$_2$H$^+$ emission in three clouds, IC 5146, (Bergin et al. 2001) B68 (Bergin et al. 2002) and FeSt 1-457 (Aguti et al. 2007). In all these clouds N$_2$H$^+$ is detected over visual extinctions of 6 magnitudes and greater, supporting the suggestion that volume densities of n(H$_2$) $\geq$ 10$^4$ cm$^{-3}$ correspond to extinctions of A$_V$ $\geq$ 6 magnitudes (and A$_K$ $\geq$ 0.7 magnitudes).  These considerations suggest that n$_0$ $\approx$ 10$^4$ cm$^{-3}$.  Although the actual value of n$_0$ is somewhat uncertain, we will from here forward assume that high extinction material corresponds to material at high volume density.

The strong and essentially linear  correlation between the SFR and cloud mass above the threshold extinction is perhaps the most interesting of our results because 1) the relatively small uncertainty in the fit suggests that this linear relation has some predictive value and 2) this suggests that the SFR in molecular gas is determined simply by the total amount of gas within the cloud that is above a certain threshold  density. If this is correct, then the key to understanding the conversion of molecular gas into stars would be to understand how high density gas is produced in clouds and what factors lead to the large variations in the relative fractions of dense gas in the present molecular cloud population.   

Establishment of the validity of the approximate linear correlation between the mass of dense gas and the star formation rate is therefore clearly important. There is well known independent evidence in both Galactic and extragalactic molecular clouds for such an intimate relation between dense molecular gas and star formation.  Gao and Solomon (2004) discovered a linear correlation between the luminosity of the HCN molecule  and the far infrared (FIR) luminosity of normal spiral  as well as starburst galaxies. As mentioned above the HCN molecule requires high densities ($>$ 10$^4$ \cc) to be excited to detectable levels and measurements of the luminosity in its lines correspond to measurements of the total amount of gas above the critical density for its excitation. The FIR luminosity is an oft used proxy for the SFR in these galaxies. Wu et al (2005) (see also  Wu et al. 2010) observed a sample of 47 massive dense cores in the Galaxy in the HCN molecule and found a linear correlation between the HCN luminosity and FIR luminosity using the same analysis methods as Gao \& Solomon. The Wu et al. (2005) results compliment our results both because they use an independent tracer (HCN) of dense material and because they observed a different, largely more distant, sample of massive clouds. Interestingly the Wu et al.  clouds fall on the same linear relation as the Gao and Solomon sources, in effect extending the Gao and Solomon relation to over nearly seven orders of magnitude in FIR luminosity/SFR. In particular,  Gao and Solomon find the relation between SFR and the mass of dense gas, $M_{dg}$, to be: SFR(\msun yr$^{-1}$) $=$ 1.8  $\times$ 10$^{-8}$ $M_{dg}$(\msun). Wu et al. derive a nearly identical result (1.2 $\times$ 10$^{-8}$ $M_{dg}$). In Figure 4 we plot the best fit linear relation to our data and from the coefficient of this fit we derive SFR (\msun\ yr$^{-1}$)  $=$ 4.6 $\pm$ 2.6$\times$ 10$^{-8}$ $M_{0.8}$ (\msun) for our data. (A similar result can be derived directly from the data in Table 2). Our results appear to be in excellent agreement with those of Gao and Solomon and Wu et al. given both our formal uncertainties and those inherent in the studies of Gao and Solomon and Wu et al. (i.e., uncertainties in the conversion of FIR luminosity into SFR, and the conversion of HCN measurements into cloud masses).

In a more recent study examining the Sptizer C2D survey of nearby dark clouds, Heiderman et al. (2010) have measured the surface densities of the youngest YSOs (Class I-type objects) and compared them to the gas mass surface densities derived from extinction and CO measurements. These measurements differ from ours in that the Heiderman et al.  measurements are of SFR surface densities (i.e., $\Sigma_{SFR}$). Thus they are not global SFR values as are our measurements. Moreover  they are spatially resolved measurements obtained across the individual clouds in their sample and they also pertain to only the youngest YSOs, those not likely to have moved too far from their birth positions.  They find strong evidence for a rapid rise of  $\Sigma_{SFR}$ at cloud gas surface densities of $\Sigma_{gas}$  $\sim$ 100 \msunsp pc$^{-2}$. At higher gas surface densities  their measurements appear to connect smoothly to those of Wu et al. (2005, 2010). Heiderman et al. have independently established the existence of a column density threshold for active star formation and, in addition, measured the location of the threshold to be approximately 100 \msunsp pc$^{-2}$, close to our  estimated value of 116 \msunsp pc$^{-2}$. We note that these results are also in agreement with the work of  Johnston, Di Francesco \& Kirk (2004) whose submillimeter survey of the Ophiuchus cloud provided evidence for an extinction threshold at about  A$_V$ = 7 magnitudes (or 112 \msunsp yr$^{-1}$) for protostellar cores in that cloud. 

The results of our paper together with previous studies such as those of Lada (1992), Lada, Evans \& Falgarone (1996), Gao \& Solomon (2004), Wu et al. (2005), Wu et al. (2010) etc. strongly indicate that it is primarily the dense gas component (i.e., n(H$_2$) $\ga$ 10$^4$ \cc) of molecular clouds that actively participates in star formation. Moreover, it now appears that the amount of star formation, measured either through the total yield of YSOs  or the SFR in a cloud depends directly and in a linear fashion on the total amount of dense gas in the cloud. Because there is a large variation in the fraction of high extinction material and presumably dense gas within individual clouds (e.g., Lombardi et al 2008, 2010; Lada et al. 2009; Figure  2 in Froebrich \& Rowles 2010), there is a large variation in the star formation rate between clouds, including those of similar mass. {\sl It is not the total mass of a cloud that controls the rate of star formation but instead the total mass of dense material in the cloud that controls the level of star formation within it}. It is not clear what determines the fraction of a molecular cloud mass that is at high density. Variations could naturally arise as a result of cloud evolution, perhaps more evolved clouds having higher fractions of their gaseous content at high, star-forming, density. Such variations could also be the result of the manner in which the cloud forms and not change substantially with time. 

It is interesting in this context to consider other recent studies of cloud structure based on extinction surveys. Surveys by Kainulainen et al. (2009), Lombardi et al. (2008; 2010) and Froebrich \& Rowles (2010) have found that the frequency distribution of extinctions in clouds can be well described by log-normal functions at low extinctions where most (90 - 99\%) of the cloud mass is found. However, significant departures from these log normal distributions are found at higher extinctions. From an analysis of 16 molecular clouds Froebrich \& Rowles (2010) found a universal threshold at A$_V$ = 6.0 $\pm$ 1.5 magnitudes, above which all the clouds in their sample showed a significant excess of material above the log-normal distribution that described the lower extinction regions of the clouds, while Kainulainen et al.(2009)  and Lombardi et al. (2010) find a somewhat lower threshold. The departure from log-normal form has been interpreted as an indication that above these threshold extinctions gravity (rather than turbulence) dominates the structure of the cloud (Lombardi et al. 2010) and that this is the material that forms stars (Kainulainen et al. 2009).  Understanding what controls the transition between turbulent and gravitationally dominated gas in a molecular cloud is clearly of interest for understanding SFR variations in molecular clouds.
  
 \section{Concluding Remarks}
  
Ever since the pioneering work of Schmidt (1959) a half-century ago there has been great interest in finding an appropriate empirical relation that would directly connect some global physical property of interstellar gas with the star formation activity in a galaxy.  Schmidt (1959) conjectured that this might take the form of a relation between  the rate of star formation and some power, n,  of the surface density of atomic (HI) gas. From evaluation of  observations available at that time he suggested that n$\approx$ 2. The discovery of molecular clouds as the sites of star formation in galaxies coupled with advances in infrared and ultraviolet observations of galaxies led to significant refinements in determinations of galaxy star formation rates and total (HI $+$ H$_2$) gas surface densities resulting in an improved determination of the relation between the surface densities of the star formation rate and the gaseous mass  and suggested that n $\approx$ 1.4 (e.g., Kennicutt 1998). 
Wong and Blitz (2002) argued that HI had little to do with star formation and obtained a similar value of n $=$ 1.4 using only the surface density of molecular hydrogen (derived from CO observations). This suggests that H$_2$ gas dominates HI gas over the range over which the measurements were carried out. As discussed earlier, Gao \& Solomon discovered a linear (i.e., n $=$ 1) correlation between SFR and HCN luminosity, which traced gas at high density. However,  Krumholtz \& Thompson (2007) and Narayanan et al. (2008) showed that when molecular-line observations are used to measure the gas surface density, the spectral index, n, derived for the empirical relationship between $\Sigma_{SFR}$  and $\Sigma_{gas}$ will depend on the specific transition or molecular-line tracer used. In particular, the slope of the empirical star formation "law" derived using molecular-line observations depends on the relation between the critical density (for excitation) of the specific tracer and the mean density of the cloudy material. In such a situation it is unclear which, if any, of the observed relations is the underlying fundamental relation. However, we would expect that using extinction to trace the gaseous component of star forming regions would not suffer from this limitation.

In this paper we systematically examined a local sample of molecular clouds using extinction measurements as a proxy to  trace their gaseous content and structure. We find that star formation is most intimately associated with the cloudy material at relatively high extinctions  and thus high volume densities. Moreover, we show that for local (d $\leq$ 500 pc) molecular clouds there exists a linear relationship between the SFR and the mass of a cloud  above a threshold extinction of A$_K \approx$ 0.8 mag (i.e., A$_V$ $\approx$ 7.3 mag, $\Sigma_{gas} \approx $ 116 \msunsp pc$^{-2}$).  The close correspondence between these results and similar ones obtained by Gao and Solomon for external galaxies suggests that this relationship is likely the underlying physical relationship or empirical law that best connects star formation activity with interstellar gas over many spatial scales within and between individual galaxies.  Therefore understanding the origin of the dense component of the molecular interstellar medium and how that component evolves may provide the key to the development of a predictive theory that links star formation and galaxy evolution.

\acknowledgments

We thank Leo Blitz,  Desika Narayanan, Phil Myers, Jan Frobrich, Neal Evans and Amanda Heiderman for informative discussions. We thank Dawn Peterson for providing  positions for the YSO population in RCrA in advance of publication. We also thank an anonymous referee for constructive criticisms that improved the paper.

\clearpage




\clearpage


\begin{figure}
\includegraphics[]{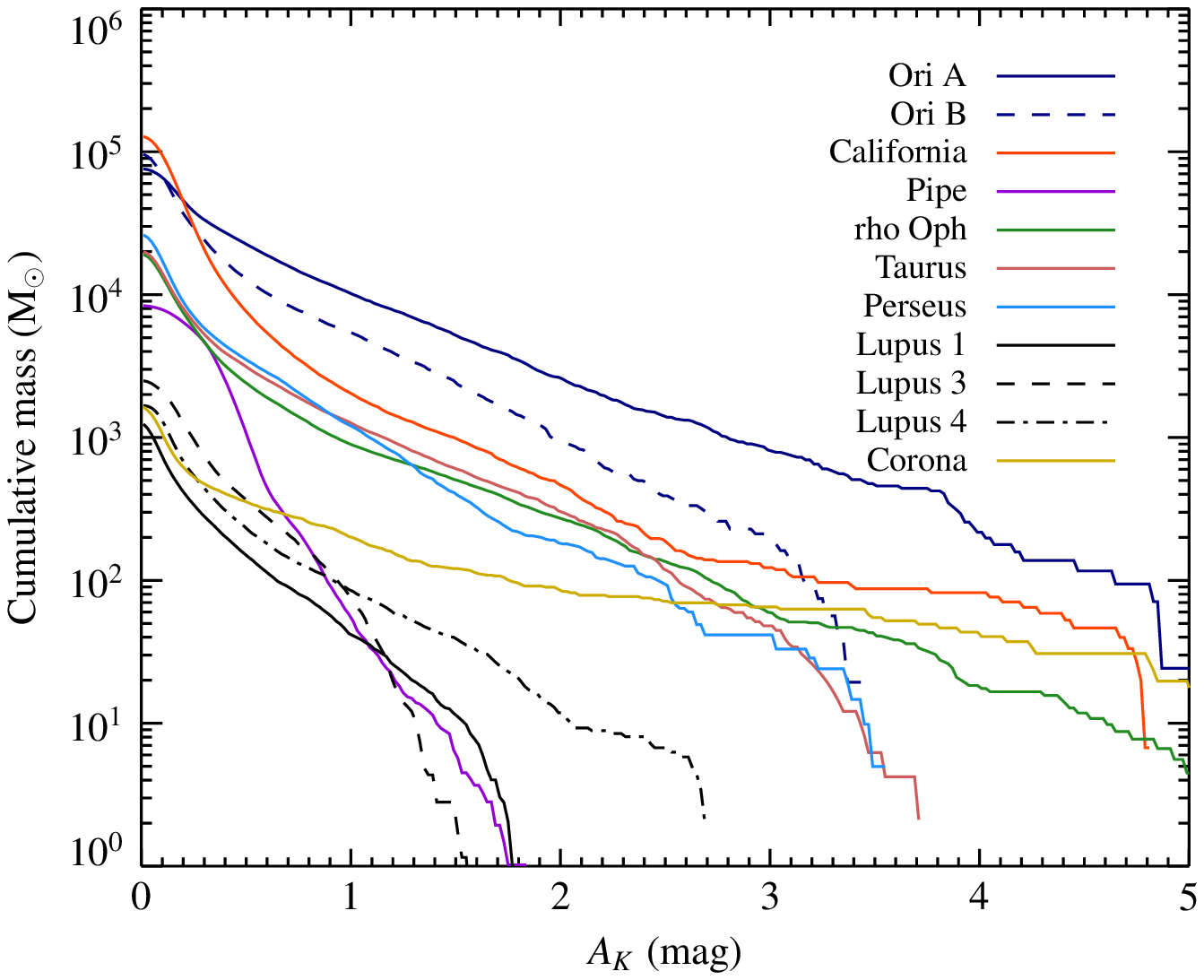}
\caption{Cumulative mass profiles as a function of infrared extinction for eleven local clouds derived from NICEST extinction maps (see text). \label{fig1}}
\end{figure}

\begin{figure}
\includegraphics[]{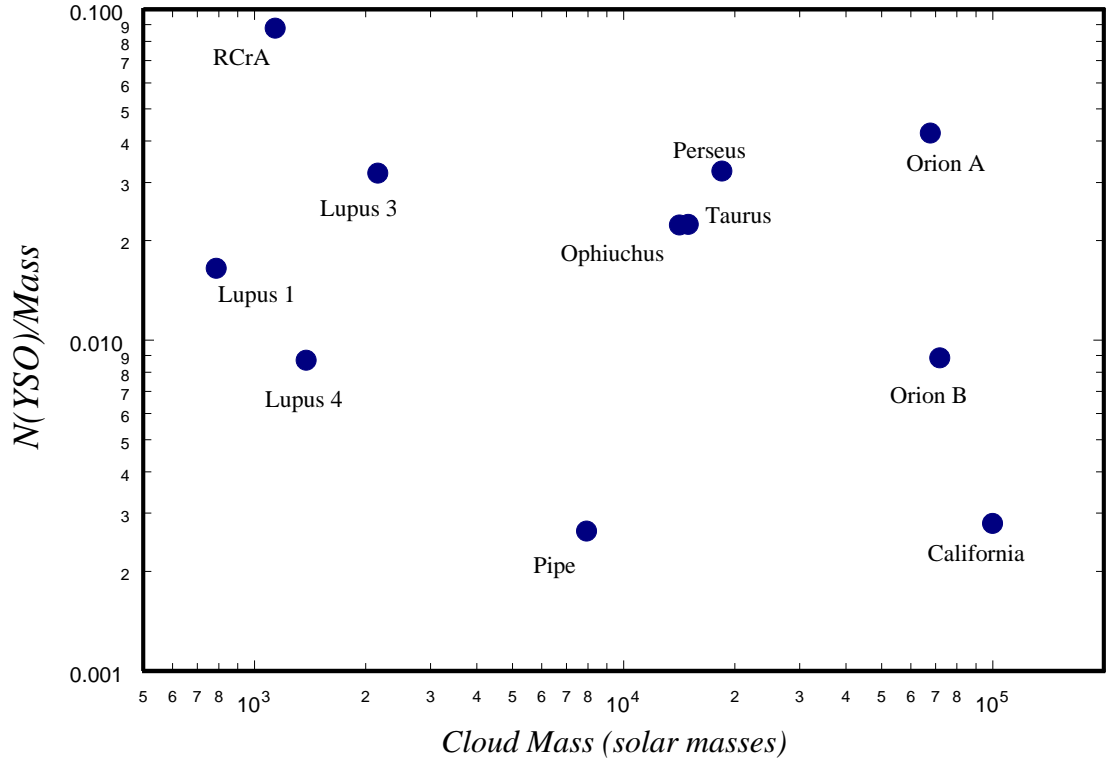}
\caption{Plot of the ratio of the total YSO content of a cloud  to the total cloud mass vs total cloud mass. This is equivalent to a measure of the star formation efficiency as a function of  cloud mass for the local sample. It is also equivalent to the measure of the star formation rate per unit cloud mass as a function of cloud mass. The plot shows large variations in the efficiency and thus the star formation rate per unit mass for the local cloud sample. (see text).}
\end{figure}

\begin{figure}
\includegraphics[]{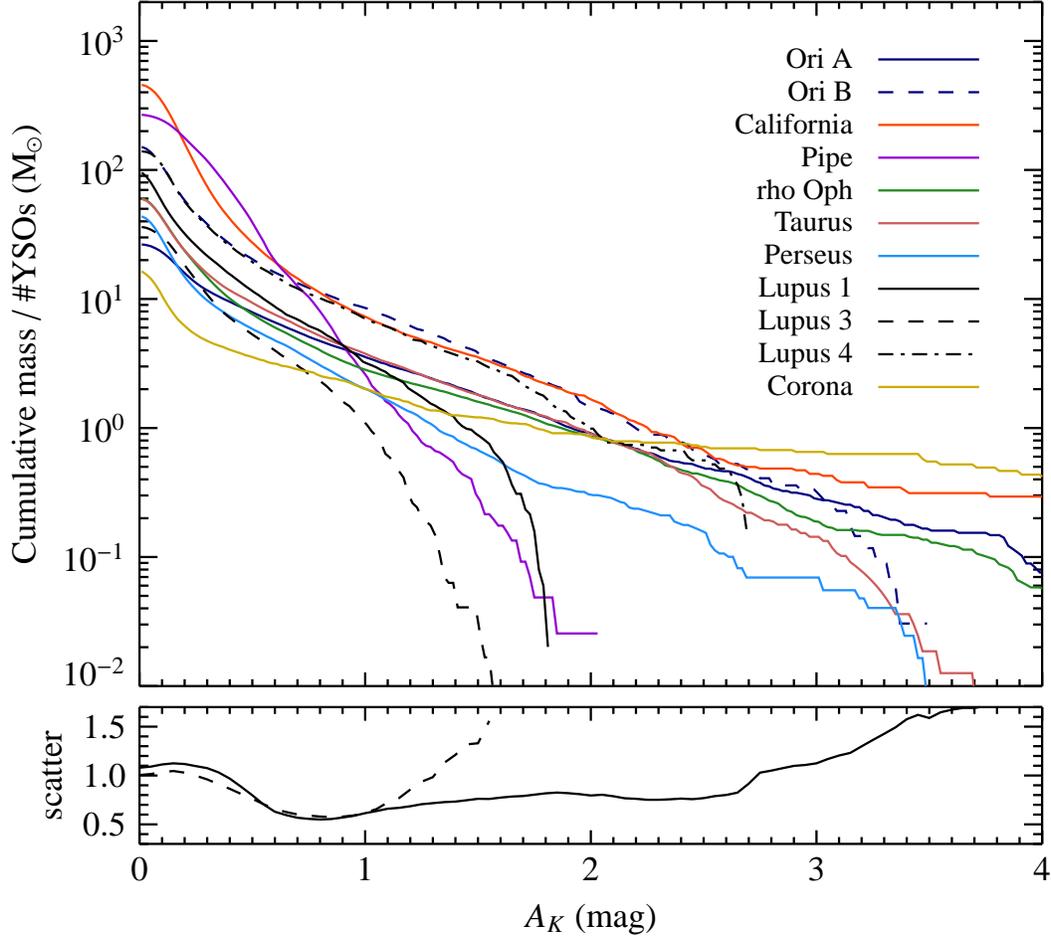}
\caption{Plot of the ratio of cumulative mass to YSO content vs extinction for the local sample of star forming molecular clouds (top). The dispersion in the logarithms of these ratios (dashed trace) and the normalized dispersion (solid trace) are also plotted as a function of extinction (bottom). The minimum in the dispersions near A$_K$ $=$ 0.8-0.9 magnitudes indicates that there is an extinction at which the total cloud mass contained above that extinction reflects most directly the star formation rates in the clouds (see text).}
\end{figure}

\begin{figure}
\includegraphics[]{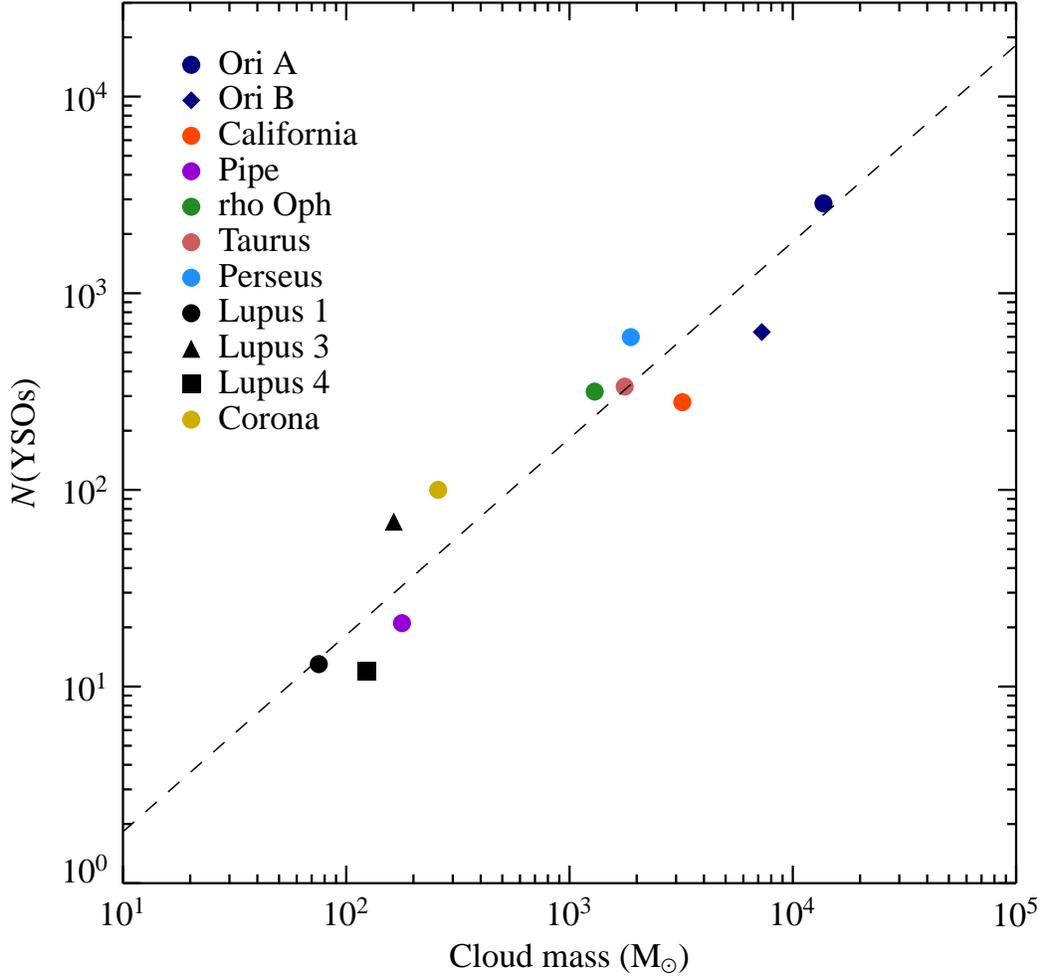}
\caption{The relation between N(YSOs), the number of YSOs in a cloud, and $M_{0.8}$,   the  integrated cloud mass above the threshold extinction of A$_{K0}$ $=$ 0.8 magnitudes. For these clouds the star formation rate (SFR) is directly proportional to N(YSOs) and thus this graph also represents  the relation between the SFR and the mass of highly extincted and dense cloud material. A line representing the best fit linear relation is also plotted for comparison.  There appears to be a strong linear  correlation between  N(YSOs) (or SFR) and $M_{0.8}$, the cloud mass at high extinction and density.}
\end{figure}







\clearpage
\begin{deluxetable}{llcl}
\tablewidth{0pt}
\tablecaption{The Local Molecular Cloud Sample}
\tablehead{
\colhead{Cloud}           & \colhead{Distance}      
& \colhead{References}  &
\colhead{Pixel Scale (deg)$^a$}    }

\startdata
Orion A&371 $\pm$ 10& 1& 0.0250\\ 
Orion B&398 $\pm$ 12& 1& 0.0250\\
California&450 $\pm$ 23& 2, 3& 0.0111\\
Perseus&240 $\pm$ 13&3& 0.0208\\
Taurus&153 $\pm$ 8& 3 & 0.0208\\
Ophiuchus&119 $\pm$ 6& 4 & 0.0166\\
RCrA&148 $\pm$30&5 & 0.0250\\
Lupus 3& 230 $\pm$ 30& 5& 0.0111\\
Lupus 4& 162 $\pm$ 30& 5& 0.0111\\
Lupus 1& 144 $\pm$ 30& 5& 0.0111\\
Pipe&130 $\pm$ 18& 6 & 0.0166\\
\enddata
\tablenotetext{a}{$=$ 0.5 $\times$ angular resolution}
\tablerefs{
(1) Lombardi, Lada \& Alves 2010 (in preparation); (2) Lada, Lombardi \& Alves 2009; (3) Lombardi, Lada \& Alves 2010;  (4) Lombardi, Lada \& Alves 2008; (5) Kunde 2010;  (6) Lombardi, Alves \& Lada 2006}
\end{deluxetable}



\clearpage
\begin{deluxetable}{lccccc}
\tablewidth{0pt}
\tablecaption{Masses and YSO Contents of Local Molecular Clouds}
\tablehead{
\colhead{Cloud}           & \colhead{Mass (\msun)$^a$}      & \colhead{Mass (\msun)$^b$}      & \colhead{No. of YSOs}          & \colhead{References}  &
\colhead{SFR(10$^{-6}\ $\msun yr$^{-1}$)}    }    

\startdata
Orion A&67,714&13,721&2862&1,2,3&715\\
Orion B&71,828&7261&635&4,5&159\\
California&99,930&3199&279&6,7&70\\
Perseus&18,438&1880&598&8,9,10&150\\
Taurus&14,964&1766&335&11&84\\
Ophiuchus&14,165&1296&316&12&79\\
RCrA&1,137&258&100&13,14, 15&25\\
Pipe&7,937$^c$&178&21&16&5\\
Lupus 3&2,157&163&69&17, 18, & 17\\
Lupus 4&1,379&124&12&17, 18, & 3\\
Lupus 1&787&75&13&17, 18, & 3\\

\enddata
\tablenotetext{a}{A$_K \geq$ 0.1 mag using NICEST.}
\tablenotetext{b}{A$_K \geq$ 0.8 mag using NICEST.}
\tablenotetext{c}{Corrected for background extinction of A$_K$ $=$ 0.15 mag.}
\tablerefs{
(1) Allen \& Davis 2008; (2) Hillenbrand 1997; (3) Peterson \& Megeath 2008;  (4) Lada et al. 1991; (5) Gib 2008;  (6) Wolk et al. 2010; (7) Lada et al. 2009; (8) Lada et al. 2006; (9) Lada et al. 1996; (10) Jorgensen, et al. 2006; (11) Kenyon et al. 2008; (12) Wilking et al. 2008;  (13) Forbrich \& Preibisch 2007; (14) Neuhauser \& Forbrich 2008; (15) Peterson et al. 2010 (16) Forbrich et al. 2009; (17) Mer{\'{\i}}n et al. 2008; (18) Comer{\'o}n 2008 }
\end{deluxetable}

\clearpage



\end{document}